\documentclass[pdftex,10pt,conference]{IEEEtran}
\usepackage[dvips]{graphicx}
\usepackage{graphics}
\usepackage{amsmath,amsfonts,amsthm}
\usepackage{cite}
\usepackage{url}
\usepackage{array}
\usepackage{makecell}
\usepackage[usenames,dvipsnames,table]{xcolor}

\title{Computation Offloading Decisions for Reducing Completion Time}

\author{
\IEEEauthorblockN{Salvador Melendez and Michael P. McGarry} 
\IEEEauthorblockA{Department of Electrical and Computer Engineering\\
University of Texas at El Paso\\
El Paso, TX 79968}
}

\begin{document}

\newcommand{\argmax}{\operatornamewithlimits{arg\,max}}
\newcommand{\argmin}{\operatornamewithlimits{arg\,min}}

\maketitle

\begin{abstract}
We analyze the conditions in which offloading computation reduces completion time. We extend the
existing literature by deriving an inequality (Eq. \ref{eq:compoff}) that relates computation offloading 
system parameters to the bits per instruction ratio of a computational job. This ratio is the inverse of 
the arithmetic intensity. We then discuss how this inequality can be used to determine the computations 
that can benefit from offloading as well as the computation offloading systems required to make offloading 
beneficial for particular computations.
\end{abstract}

\begin{keywords}
Cloud computing, mobile cloud computing, communication, networks, cloud resource provisioning.
\end{keywords}

\section{Introduction}
\label{sec:intro}
Computing as a utility offers computation as a service over a communication network in much the same way that the electrical
utility offers electric power as a service over a power distribution network. Several advances have occurred in recent years 
to make computing as a utility commercially viable. These include advances in computing resource virtualization/isolation, 
advances in high-bandwidth/low-latency communication, and increasing economies of scale for large-scale computing facilities. 
Cloud computing~\cite{DKMPV0909,AFGJKKLPRSZ0410} is a recent buzz word that encompasses computing as a utility. Many computing 
scenarios benefit from computing as a utility, such as those where the computing demand is elastic or unpredictable. However, 
one of the most compelling uses of computing as a utility is to enhance the capabilities of edge computing 
devices~\cite{SBCD1009,CBCWSCB0610} such as smart phones, tablets, wearable computers, smart objects (e.g., smart appliances 
and furniture), and cyber-physical systems (CPS). Edge computing devices can be empowered by computing as a utility, through 
remote execution, to execute real-time applications that would not be possible on the edge device itself in the desired 
time-frame. The act of remote execution on a so called ``cloud resource'' is often referred to as 
\textbf{computation offloading}~\cite{PM1083} or cyber-foraging~\cite{BFSSY0902,KB1210,SKK1012}.

\begin{figure*}
	\centering
	\includegraphics[scale=0.35]{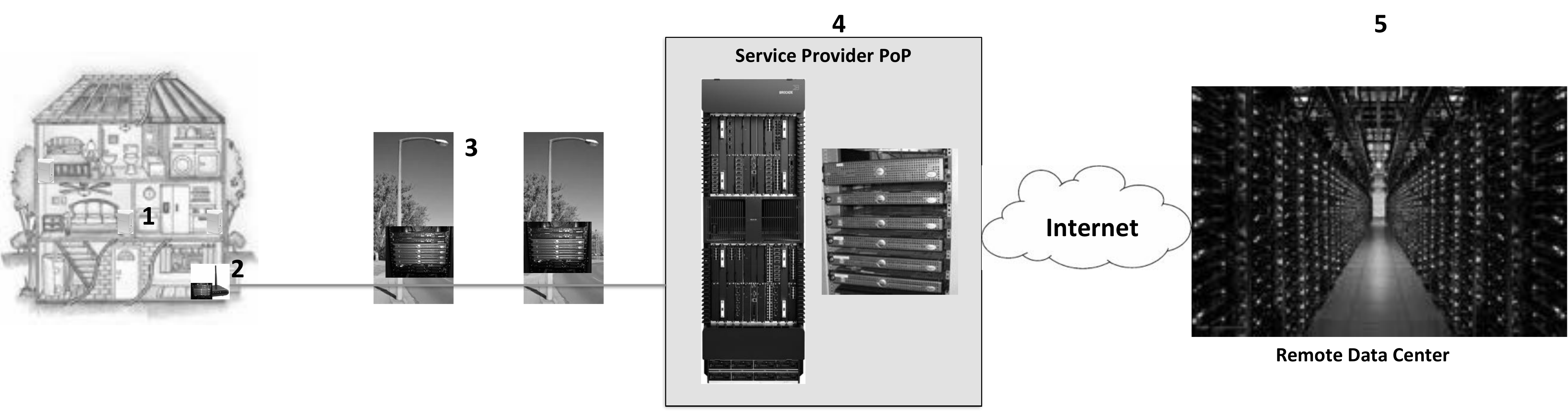}
	\caption{A five tier computation offloading system: ($1$) ``plug computers'' in certain rooms of a home (dual-core ARM systems) plugged into
	power outlets, ($2$) server access points that are a combined wireless access point and multiprocessor/multi-core computer, ($3$) small racks
	of computers attached to light posts in the community, ($4$) large racks of computers alongside racks of telecommunications switching equipment
	in telecom points-of-presence, ($5$) remote data center (e.g., Amazon Web Services).}
	\label{fig:offloadfivetier}
\end{figure*}

The computing devices that provide the computation offloading service to edge computing devices can exist not only in a remote 
data center but in locations much closer to the edge. These locations may be within the same room, at an Internet access
point, or inside an Internet Service Provider (ISP) point-of-presence (PoP). These various locations provide a tiered
computing structure, see Figure \ref{fig:offloadfivetier}, whereby cloud computing resources can be found at varying distances
from the edge computing devices.

Recent research activities have produced a variety of programming frameworks and systems that make computation offloading 
possible. Spectra~\cite{FPS0702} and Chroma~\cite{BSPO0503} were seminal computation offloading systems that were followed up 
by several others (e.g., Slingshot~\cite{SF0605}, MAUI~\cite{CBCWSCB0610}, Cuckoo~\cite{KPKB1010}, CloneCloud~\cite{CIMNP0411}). 
Clearly, mechanisms to facilitate computation offloading have received significant attention from the research community. A recent 
survey on computation offloading~\cite{KLLB0213} identified two objectives for computation offloading: ($1$) reduce execution time, 
and ($2$) shift energy consumption. The survey also identified two classes of decision to be made: ($1$) \textit{what} computation 
to offload, and ($2$) \textit{where} to offload computation.

The decision regarding \textit{what} to offload is generally referred to as the partitioning problem and many techniques 
have been proposed and evaluated~\cite{LWX1093,LWX1101,LX1102,GNMGM0303,CSWKK0104,WL0604,WL0604b,OYL0306,OYLH0607,HKL0509,
CM0610,HZY0710,NKLL1010,SK0511,HWN0612,ZLJF1112,YCYLHC0313}. Applications are partitioned into components and a binary decision 
is made whether to offload a component or not. The data exchanged between the application components is considered when making 
the decision to offload each component.

Much of the literature related to deciding \textit{where} to offload focuses on the binary decision of whether to offload or
not, which is similar to the decision of \textit{what} to offload. Some of the analysis developed is static and provides
rules of thumb regarding the conditions in which offloading computation is favorable~\cite{KL0410,WWW0613}. The general 
consensus is that it is beneficial to offload computation if there are large amounts of computation and only a small amount of 
data that needs to be transmitted over a communication network. Most of the systems mentioned above that facilitate computation
offloading (e.g., MAUI, Cuckoo) indeed included a computation offloading decision algorithm that decided whether to offload a 
computation or not (i.e., the binary decision). For example, MAUI~\cite{CBCWSCB0610} contains a binary decision algorithm that 
is the solution of a binary integer linear mathematical program with an objective function to minimize energy consumption subject 
to a particular completion time constraint. The decision algorithm utilizes historical energy consumption and network throughput 
data and executes on the offload target to avoid burdening the mobile device.

In this paper we extend the literature that analyzes when offloading reduces completion time. We derive an inequality that 
compares offloading system parameters to computational job parameters to determine when offloading would reduce completion time. 
The computational job parameters we use form a ratio that is related to the arithmetic intensity of a computation~\cite{WWP0409}.

\subsection{Overview}
In Section \ref{sec:model} we present the system model that we use in our subsequent analysis. In Section \ref{sec:offload}
we present our analysis of the conditions in which computation offloading can reduce execution time. Finally, in Section 
\ref{sec:conclusion} we conclude with a summary of our findings.

\section{System model}
\label{sec:model}
The system under study consists of a client device that produces computational jobs that can either be executed locally or 
offloaded for execution on one of several cloud resources. The cloud resources are distributed spatially in a network at 
varying numbers of hops from the client device. If the job is executed locally, the completion time is the time to execute 
all of the instructions of the job at the execution rate of the local computing device (i.e., computation time). If the job 
is offloaded, the completion time is the time to execute all of the instructions of the job at the execution rate of the 
selected cloud resource (i.e., computation time) plus the time to transmit the input and output data through the network 
(i.e., communication time). In our analysis, we consider the memory access times to be similar between the local computing 
device and the cloud resources. This is a reasonable assumption given the well documented processor-memory performance 
gap~\cite{comparchbook}. We now present our models for computation time and communication time that compose our model of 
completion time.

\subsection{Computation time}
Let $C$ be the size of the computational job (instructions), $e$ be the execution rate of the local computing device 
(instructions/sec), and $E$ be the execution rate of that particular compute resource (instructions/sec). The computation 
time for the computational job executed locally is $\xi = \frac{C}{e}$ and $\xi = \frac{C}{E}$ if it is executed on a
particular cloud resource.

\subsection{Communication time}
We use a per-hop model of communication time that incorporates the generally accepted categories of delay incurred at 
each hop~\cite{netbook}: ($i$) processing, ($ii$) queueing, ($iii$) transmission, and ($iv$) propagation delays.
Let $S$ be the size of the packet (bits), $\alpha$ be the 
processing delay at a hop in a network (sec), $\beta$ be the queueing delay incurred at that hop (sec), $\gamma$ be the rate of 
the transmission channel at that hop (bits/sec), $l$ be the length of that hop (meters), and $c$ be the speed of light (meters/sec). 
Then, the time required to transmit a packet across one hop in the network is 
$\psi = \alpha + \beta + \frac{S}{\gamma} + \frac{l}{\frac{2}{3}c}$.

Let $h$ be the number of hops along the path that a packet traverses from its source to its destination. Ignoring the 
negligible processing and propagation delays, we transform the equation above to characterize the end-to-end communication 
time as $\psi = \sum_{j=1}^{h} \left ( \beta(j) + \frac{S}{\gamma(j)} \right)$.

Now, consider the transmission of a file that is larger than the maximum allowable packet size. In this case, the file is 
transformed into a packet train: several maximum-size packets followed by a remainder packet that can be up to the maximum 
size allowed. As individual packets in the packet train are transmitted through several hops in the network, packet 
transmission can occur in parallel. To model the communication time considering the effect of parallel packet transmission, 
we consider: ($1$) the time to transmit the entire file through the bottleneck transmission channel, and ($2$) the time to 
transmit the last packet in the packet train through each hop. Let $F$ be the size of the file (bits) and $N$ be the size of 
the last packet in the packet train. Then the communication time for the entire file is

\begin{equation*}
\psi = \frac{F}{\min\{\gamma(j), \forall_{j}\}} + \sum_{j=1}^{h} \left ( \beta(j) + \frac{N}{\gamma(j)} \right) \mbox{.}
\end{equation*}

\subsection{Composite completion time model}
Let $x$ be the completion time, $I$ be the size of the input data (bits), and $O$ be the size of the output data (bits). 
If the computational job is executed locally, the completion time is:

\begin{equation*}
x = \frac{C}{e} \mbox{.}
\end{equation*}

\noindent If the computational job is executed on a particular cloud resource, the completion time is:

\begin{equation*}
\frac{C}{E} + \frac{(I+O)}{\min\{\gamma(j),\forall_{j}\}} + \sum_{j=1}^{h} \left ( \beta(j) + \frac{N}{\gamma(j)} \right) \mbox{.}
\end{equation*}

\subsection{Useful ratios}
The following two ratios are useful for our computation offloading analysis.

The computing-to-communication ratio (CCR) is the ratio of the computation time to the communication time. Using the
symbols above,

\begin{equation*}
CCR = \frac{\xi}{\psi} \mbox{.}
\end{equation*}

The remote-to-local ratio (RLR) is the ratio of the cloud resource execution speed to the local execution speed. Again,
using the symbols above,

\begin{equation*}
RLR = \frac{E}{e} \mbox{.}
\end{equation*}

\section{When to offload computation?}
\label{sec:offload}
To favor remote execution (or computation offloading) for reducing completion time, the following inequality must hold true:

\begin{equation}
\frac{C}{e} > \left[ \frac{C}{E} + \frac{(I+O)}{\min\{\gamma(j),\forall_{j}\}} + \sum_{j=1}^{h} \left ( \beta(j) + \frac{N}{\gamma(j)} \right) \right]\mbox{.}
\end{equation}

We now manipulate this inequality to derive useful insight into computation offloading system design. To ease manipulation
of this inequality, we let $H$ represent the hop-by-hop network delay that is agnostic to the job size or the data size, 
$F$ be all of the data to be transferred over the network (input and output), and $\Gamma$ represent the transmission 
rate of the bottleneck link in the network. We now isolate the RLR on the left hand side of the inequality.

\begin{equation*}
\frac{C}{e} > \left[ \frac{C}{E} + \left( \frac{F}{\Gamma} + H \right) \right]
\end{equation*}

\begin{equation}
C \left(\frac{E}{e} - 1\right) > E\left( \frac{F}{\Gamma} + H \right)
\label{eq:ineq1}
\end{equation}

\begin{equation*}
\frac{E}{e} > \frac{\left( \frac{F}{\Gamma} + H \right)}{\frac{C}{E}} + 1; \frac{E}{e} > \frac{\psi}{\xi} + 1
\end{equation*}

\begin{equation}
RLR > \frac{1}{CCR} + 1
\label{eq:rlrccr}
\end{equation}

After isolating the RLR on the left hand side, we have the inequality shown in Eq. \ref{eq:rlrccr}.
That inequality shows there is a nearly inverse relationship between RLR and CCR. To visualize the
implications of this inequality we tabulate the RLR value required to make offloading favorable for
various values of the CCR; see Table \ref{tab:rlrccr}. A CCR of $1x10^{-3}$ requires cloud resources 
to be more than a thousand times faster than the local computing device. A CCR of $1x10^{-6}$ requires the 
cloud resources to be a million times faster! To obtain a sense of practical RLR values, we have 
compiled the instructions per second (IPS) ratings of two embedded processors that represent the low 
end (Texas Instrument's MSP430) and the high end (Apple's A9) of the spectrum of the embedded
processors deployed in handheld devices. To complete the RLR computation, we have compiled the IPS 
ratings of a laptop-class processor (Intel's Celeron), desktop-class processor (Intel's Core i3), 
and a server-class processor (Intel's Xeon). See Table \ref{tab:proc} for these IPS ratings.

\begin{table}
\caption{RLR values required for offloading to be favorable for several CCR values.}
\begin{center}
\begin{tabular}{| c || c |}
	\hline
	\textbf{CCR} & \textbf{RLR} \\
	\hline \hline
	1e-6 & $\approx$ 1e6 \\
	\hline
	1e-3 & 1001 \\
	\hline
	0.01 & 101 \\
	\hline
	0.1 & 11 \\
	\hline
	1 & 2 \\
	\hline
	1e3 & 1.001 \\
	\hline
	1e6 & 1.000001 \\
	\hline
\end{tabular}
\label{tab:rlrccr}
\end{center}
\end{table}

\begin{table}
\caption{Instructions per second ratings for several processors.}
\begin{center}
\begin{tabular}{| c || c |}
	\hline
	\textbf{Processor} & \textbf{IPS} \\
	\hline \hline
	MSP430 & 16x10\textsuperscript{6}~\cite{MSP430}\\
	\hline
	A9 & 3.6x10\textsuperscript{9}~\cite{A9} \\
	\hline
	Celeron & 6.43x10\textsuperscript{9}~\cite{celeron_xeon} \\
	\hline
	Core i3 & 36.8x10\textsuperscript{9}~\cite{i3} \\
	\hline
	Xeon & 136.20x10\textsuperscript{9}~\cite{celeron_xeon} \\
	\hline
\end{tabular}
\label{tab:proc}
\end{center}
\end{table}

Using the IPS ratings shown in Table \ref{tab:proc} we compute the RLR value for each pair of
handheld and laptop/desktop/server class processor. These RLR values are shown in Table \ref{tab:rlr}.
We see the RLR values range from 1.79 for an A9 processor offloading to a Celeron processor (CCR must be
greater than 1.27 to offload) up to 8512.5 for an MSP 430 processor offloading to a Xeon processor (CCR 
must be greater than 1.1x10\textsuperscript{-4} to offload).

\begin{table}
\caption{CPU RLR values}
\begin{center}
\begin{tabular}{|l|c|c|c|c|}
	\hline
\diaghead{\theadfont Diag IIIIIIIIIIIII}%
{Local}{Remote}&
\thead{\textbf{Celeron}}&\thead{\textbf{i3}}&\thead{\textbf{Xeon}}\\
    \hline
\textbf{MSP430} & 401.875 & 2300 & 8512.5\\
    \hline
\textbf{A9} & 1.78611 & 10.222 & 37.833\\
    \hline
\end{tabular}
\label{tab:rlr}
\end{center}
\end{table}

Equation \ref{eq:rlrccr} leads one to, at first glance, think that if we increase the RLR value by let's
say increasing the execution speed of the remote resource (i.e., increase $E$) this will make computation
offloading beneficial for more applications as characterized by their CCR value. However, if we look back 
at the presentation of this inequality specifically in Eq. \ref{eq:ineq1} we see that increasing $E$ will
increase both sides of the inequality. Therefore, increasing $E$ does not make offloading favorable for more
applications.

To gain clear insight into how system parameters should be manipulated to increase the number of applications 
that benefit from computation offloading we need an inequality with job parameters on one side of the inequality
and computation offloading system parameters on the other side of the inequality.

Starting from the original simplified inequality in Eq. \ref{eq:ineq1} we take steps to move the offloading system 
parameters to the left hand side.

\begin{equation*}
C \left(\frac{1}{e} - \frac{1}{E}\right) > \left( \frac{F}{\Gamma} + H \right); C\Gamma \left(\frac{1}{e} - \frac{1}{E}\right) > \left( F + H\Gamma \right)
\end{equation*}

\begin{equation*}
\Gamma \left(\frac{1}{e} - \frac{1}{E}\right) > \left( \frac{F}{C} + \frac{H\Gamma}{C} \right); \Gamma \left(\frac{1}{e} - \frac{1}{E} - \frac{H}{C}\right) > \frac{F}{C}
\end{equation*}

If we consider an uncongested network so that $H$ is negligible,

\begin{equation}
\Gamma \left(\frac{1}{e} - \frac{1}{E}\right) > \frac{F}{C} \mbox{.}
\label{eq:compoff}
\end{equation}

The ratio of job parameters on the right hand side of this inequality is the inverse of the arithmetic intensity
or the bits per instruction of the computational job. This context differs from the common usage of the term arithmetic 
intensity in that we are referring to bits that would be communicated over the network and not bits that would be 
read/written from/to the memory system. The left hand side represents the capacity of the offloading system to consume 
bits and instructions; at the bottleneck link rate $\Gamma$ and the cloud resource execution rate $E$ compared to the 
local execution rate $e$, respectively.

\subsection{$\frac{F}{C}$ values from TACC workload data}
We obtained computational workload data from the Texas Advanced Computing Center (TACC) in Austin, TX. The data 
was provided by the Director of High Performance Computing, Dr. Bill Barth. The trace contains records of 78,176 
computation jobs of 1,159 different applications. It provides information such as: 1) job ID, 2) application name, 
3) job size, 4) bytes written, 5) bytes read, and 6) execution time. We use this data along with some assumptions
about the execution speed of TACC resources to derive the $\frac{F}{C}$ for recognizable applications. Table 
\ref{tab:bpi} shows the minimum, average, and maximum of the derived values of each data point in the workload data
set for recognizable applications.

\begin{table}
\caption{bits per instruction for several TACC workload applications.}
\begin{center}
\begin{tabular}{| c || c | c | c |}
	\hline
	\textbf{App Name} & \textbf{min} & \textbf{ave} & \textbf{max} \\
	\hline \hline
	namd2 & 1.61x10\textsuperscript{-7} & 4.45x10\textsuperscript{-5} & 1.01x10\textsuperscript{-3}\\
	\hline
	lmp\_stampede & 1.36x10\textsuperscript{-7} & 6.60x10\textsuperscript{-6} & 9.53x10\textsuperscript{-4}\\
	\hline
	vasp\_ncl & 1.43x10\textsuperscript{-7} & 6.56x10\textsuperscript{-6} & 2.86x10\textsuperscript{-4}\\
	\hline
	cosmomc & 2.34x10\textsuperscript{-7} & 5.61x10\textsuperscript{-6} & 4.84x10\textsuperscript{-4}\\
	\hline
	charmm & 1.42x10\textsuperscript{-7} & 9.32x10\textsuperscript{-6} & 7.34x10\textsuperscript{-5}\\
	\hline
	nwchem & 1.01x10\textsuperscript{-6} & 7.06x10\textsuperscript{-5} & 1.80x10\textsuperscript{-4}\\
	\hline
	fvcom & 3.36x10\textsuperscript{-6} & 2.17x10\textsuperscript{-4} & 2.27x10\textsuperscript{-3}\\
	\hline
	mdrun\_mpi & 1.66x10\textsuperscript{-7} & 8.18x10\textsuperscript{-6} & 1.08x10\textsuperscript{-4}\\
	\hline
	siesta & 1.47x10\textsuperscript{-7} & 9.81x10\textsuperscript{-6} & 5.29x10\textsuperscript{-5}\\
	\hline
\end{tabular}
\label{tab:bpi}
\end{center}
\end{table}

\subsection{Determining when offloading is beneficial}
\label{sec:compoffanalysis}
We can use the inequality of Eq. \ref{eq:compoff} along with the inverse arithmetic intensity values (i.e., $\frac{F}{C}$)
to determine when offloading is beneficial. To assist in this comparison we compute the values of $\left(\frac{1}{e} - \frac{1}{E}\right)$
for various combinations of local and remote processors. See Table \ref{tab:rlr2} for these values.

Using this data we can see that if we are offloading from an MSP430 processor to an Intel Celeron processor through a network
with a 1 Kbps bottleneck link rate, applications with inverse arithmetic intensities less than 6.23x10\textsuperscript{-5} will 
benefit from offloading with respect to a reduction in completion time. Looking at Table \ref{tab:bpi} we see that a large number 
of the scientific applications from the TACC workload data will benefit from offloading in this scenario. If we increase the bottleneck 
link rate to 1 Mbps, then inverse arithmetic intensities less than 6.23x10\textsuperscript{-2} benefit from offloading; now all of the 
scientific applications listed in Table \ref{tab:bpi} benefit from offloading.

If we offload from Apple's A9 to an Intel Celeron through a network with a 1 Kbps bottleneck link rate then applications with inverse
arithmetic intensities less than 1.22x10\textsuperscript{-7} benefit from offloading. If the bottleneck link rate is increased to
1 Mbps, then inverse arithmetic intensities less than 1.22x10\textsuperscript{-4} benefit from offloading.

This style of analysis using our inequality from Eq. \ref{eq:compoff} can be utilized to determine which computations can benefit from
a particular offloading system or what offloading system is required to make offloading beneficial for a particular computation.

\begin{table}
\caption{Remote to local execution ratio $\left(\frac{1}{e} - \frac{1}{E}\right)$}
\begin{center}
\begin{tabular}{|l|c|c|c|c|}
	\hline
\diaghead{\theadfont Diag IIIIIIIIIIIII}%
{Local}{Remote}&
\thead{\textbf{Celeron}}&\thead{\textbf{i3}}&\thead{\textbf{Xeon}}\\
    \hline
\textbf{MSP430} & $6.23x10^{-8}$ & $6.25x10^{-8}$ & $6.25x10^{-8}$\\
    \hline
\textbf{A9} & $1.22x10^{-10}$ & $2.51x10^{-10}$ & $2.70x10^{-10}$\\
    \hline
\end{tabular}
\label{tab:rlr2}
\end{center}
\end{table}

\section{Conclusion}
\label{sec:conclusion}
We have derived an inequality that relates computation offloading system parameters to the arithmetic intensity of a computation
(see Eq. \ref{eq:compoff}). This inequality can be used to determine which computations benefit from computation offloading w.r.t. 
reduction in completion time for a particular computation offloading system. This inequality can also be used to determine the 
computation offloading system required to permit certain computations to benefit from offloading. See Section \ref{sec:compoffanalysis} 
for some examples of this type of analysis using our inequality.

Future work should tabulate the arithmetic intensities for various computations and use these values to identify which computations
can benefit from practical computation offloading systems.

\section*{Acknowledgment}
This research was supported by the U.S. Army Research Laboratory (USARL) via the Army High Performance Computing Research Center 
(AHPCRC) through Stanford University, award 60300261-10737-B. 

\bibliographystyle{IEEEtran}
\bibliography{cloud}

\end{document}